\documentclass[11pt]{article}
\usepackage{amsmath,amssymb,amsthm,mathrsfs,stackrel}
\usepackage{array}
\usepackage{graphicx,psfrag,epsf}
\usepackage{multirow}
\usepackage{multicol}
\usepackage{threeparttable}
\usepackage{booktabs}
\usepackage[round]{natbib}
\usepackage{float}
\usepackage{algorithmic}
\usepackage[ruled,vlined]{algorithm2e}
\usepackage{url}
\setcitestyle{authoryear,round}

\makeatletter
\renewcommand{\algocf@captiontext}[2]{#1\algocf@typo. \AlCapFnt{}#2}

\def\@algocf@capt@plain{top}
\renewcommand{\algocf@makecaption}[2]{%
  \addtolength{\hsize}{\algomargin}%
  \sbox\@tempboxa{\algocf@captiontext{#1}{#2}}%
  \ifdim\wd\@tempboxa >\hsize%
    \hskip .5\algomargin%
    \parbox[t]{\hsize}{\algocf@captiontext{#1}{#2}}%
  \else%
    \global\@minipagefalse%
    \hbox to\hsize{\box\@tempboxa}%
  \fi%
  \addtolength{\hsize}{-\algomargin}%
}
\makeatother

\usepackage{amsfonts,bm,amsbsy,bbm}
\usepackage{rotating,lscape}
\usepackage{booktabs}
\usepackage{xcolor}
\usepackage{xr-hyper}
\usepackage[colorlinks,citecolor=blue,linkcolor=blue,urlcolor=blue]{hyperref}

\newtheorem{theorem}{Theorem}

\newtheorem{corollary}{Corollary}

\usepackage{siunitx}             
\newcommand{\cN}{\mathcal{N}}
\providecommand{\noopsort}[1]{} 

\newcommand{\w}{\mathbbm{w}}
\newcommand{\E}{\mathbb{E}}

\newcommand{\R}{\mathbb{R}}
\def\t{\top}

\def\l{\left}
\def\r{\right}

\begin{document}

\title{Doubly robust target inference for generalized linear regression with completely missing covariates}

\author{
Huali Zhao\textsuperscript{1} and 
Ke Deng\textsuperscript{2}\thanks{kdeng@tsinghua.edu.cn}\\[0.5em]
\textsuperscript{1}School of Mathematics and Statistics, Huazhong University of Science and Technology\\
\textsuperscript{2}Department of Statistics and Data Science, Tsinghua University
}

\date{}

\maketitle
\thispagestyle{empty}
\baselineskip=20pt

\begin{abstract}
Large-scale multipurpose cohort studies and biobanks often omit covariates needed for specific downstream analyses. We study target-population inference for generalized linear regression when key covariates are completely absent from the target data but observed in a related source population. Standard missing covariate methods are not directly applicable because they require at least partial observation of the covariates in the target population. We develop a doubly robust transfer learning framework under a sub-population shift assumption, which allows the distribution of the observed outcome and covariates to differ between populations while requiring the conditional distribution of the missing covariates given the observed variables to be shared. Unlike linear regression, nonlinear target estimating equations require parameter-indexed conditional functionals of the missing covariates rather than a fixed collection of low-order conditional moments. Our estimator combines importance-weighted source estimating equations with imputation of these conditional functionals. Under the identifying assumption, the estimator remains consistent when either the density-ratio model or the conditional-covariate model is correctly specified. Under regularity conditions, it is root-$n$ consistent and asymptotically normal, and attains the semiparametric efficiency bound when both nuisance models are correctly specified.
\end{abstract}


\pagestyle{plain}

\begingroup
\allowdisplaybreaks

%%%%%%%%%%%%%%%%%%%%%%%%%%%%%%%%%%%%
\section{Introduction}
%%%%%%%%%%%%%%%%%%%%%%%%%%%%%%%%%%%%

Modern large-scale biomedical programs such as All of Us \citep{denny2019all} and the UK Biobank \citep{conroy2023uk} provide exciting new opportunities for biomedical research. Unlike traditional studies designed for specific scientific questions, these programs generate expansive multipurpose data resources aiming for broad applicability, inspiring many unexpected research interests, which creates inevitable trade-offs in variable availability. Although such resources contain rich genetic, lifestyle, clinical, and wearable-device information, they cannot comprehensively measure every covariate required for all downstream analyses. For example, the All of Us data (\href{https://allofus.nih.gov/}{allofus.nih.gov}) incorporate detailed Fitbit-derived activity data but lack comprehensive dietary information, such as fat intake, which constrains nutrition epidemiological research. This creates a cross-population missing data problem in which some covariates are completely unobserved in the target population but available in a related source population.

We study statistical inference for a target-population regression parameter defined through a generalized linear estimating equation when a block of covariates is completely missing in the target data. Let \(X\) denote the covariates absent from the target population and let \(Z\) denote covariates observed in both the target and source populations. The inferential goal is not to predict \(X\) itself, but to estimate the target-population parameter characterizing the working regression relationship between \(Y\) and \((X,Z)\). This setting differs from standard single-sample missing-covariate problems, where the covariates of interest are at least partially observed in the analysis population. Many standard estimating-equation, likelihood, multiple-imputation, and doubly robust methods rely on such partial observability and therefore are not directly applicable when \(X\) is absent for every target observation \citep{zhao1996regression,lipsitz1999weighted,bang2005doubly,wang2009statistical,kluger2025prediction}.

Transfer learning offers a natural perspective, because a source population contains complete observations on \((Y,X,Z)\). Existing closely related transfer learning methods, however, mainly address settings in which the outcome or label is completely missing in the target population. Under covariate shift, one typically assumes an invariant conditional outcome distribution and estimates a density ratio using covariates observed in both populations \citep{liu2023augmented,zhou2024model,zhou2025doubly}. This strategy breaks down for completely missing covariates, because a full covariate density ratio involving \(X\) is not estimable from the target data. Under label shift, one assumes invariance of the full covariate distribution given the outcome, such as \((X,Z)\mid Y\) \citep{lee2025doubly}, which can be restrictive in heterogeneous biomedical studies because the observed covariates \(Z\) may themselves encode demographic, geographic, clinical, or design-related heterogeneity. 

Several recent works have addressed completely missing covariates in the target population for linear regression models. \citet{zhao2026doubly} consider the special case where the missing covariates are binary, while \citet{zhao2026augmented} study more general missing covariates. In particular, \citet{zhao2026augmented} formulate a \emph{sub-population shift assumption}, under which the conditional distribution of the target-missing covariates given the observed variables is invariant across the source and target populations. We adopt the same identifying condition in this paper, but address the additional complications arising from nonlinear target estimating equations. For generalized linear estimating equations with a nonlinear inverse link, the target moment condition involves nonlinear functions of the missing covariates and the unknown regression parameter. Consequently, imputing \(X\), or even imputing finitely many low-order moments of \(X\), is generally insufficient. Instead, one must handle conditional functionals such as $\E\{g(X^\top\beta_x+Z^\top\beta_z)\mid Y,Z\}$ and $\E\{Xg(X^\top\beta_x+Z^\top\beta_z)\mid Y,Z\}$, which depend on the unknown parameter \(\beta\). This nonlinear dependence creates new challenges for constructing doubly robust estimating equations and for deriving valid inferential theory.

To this end, we propose a Doubly Robust Transfer Learning framework for target generalized linear regression with \emph{completely missing covariates} (CMC), denoted as ``DRTL-CMC''. 
Our main contributions are as follows:
\begin{itemize}    
    \item We extend target inference with completely missing covariates from linear regression to generalized linear estimating equations, where nonlinear target scores require parameter-indexed conditional functionals of the missing covariates. We develop a corresponding augmentation strategy based on these functionals.
    
    \item We establish double robustness, root-\(n\) consistency, and asymptotic normality while accounting for nuisance estimation. When both nuisance models are correctly specified, we further characterize the efficient influence function and establish semiparametric efficiency.
  
    \item We provide practical implementation strategies, including parametric and Monte Carlo evaluation of the conditional functionals, and demonstrate the method through simulations and analyses on UK Biobank data and NHANES data.
\end{itemize}

%%%%%%%%%%%%%%%%%%%%%%%%%%%%%%%%%%%%
\section{Methodology}\label{sec:method}
%%%%%%%%%%%%%%%%%%%%%%%%%%%%%%%%%%%%
%%%%%%%%%%%%%%%%%%%%%%%%
\subsection{Problem Statement and Identification Assumption}
%%%%%%%%%%%%%%%%%%%%%%%%
In this study, we consider target-population inference for a generalized linear regression parameter under the CMC scenario, with the support of a heterogeneous source population with fully observed variables. The datasets from the source population and the target population are referred to as the \emph{source data} and the \emph{target data}, respectively. The source data, indexed by $s=1$, consist of $n$ independent and identically distributed (i.i.d.) observations $\{(Y_i,X_i,Z_i)\}_{1\leq i\leq n}$ about an outcome variable $Y$ and two sets of covariates $X=(X_1,\ldots,X_p)^\t$ and $Z = (Z_1,\ldots,Z_q)^\t$ with $Z_1=1$. The target data, indexed by $s=0$, consist of $N$ i.i.d. observations $\{(Y_i,Z_i)\}_{n+1\leq i \leq n+N}$ about $Y$ and $Z$ only, with $X$ completely missing. The two populations, in general, have heterogeneous data distributions. 

Without an identifying condition linking the source and target populations, the target parameter cannot in general be recovered from the observed data. Following \citet{zhao2026augmented}, we adopt the \emph{sub-population shift assumption}. Let $p_s(y,x,z)$ denote the joint density of $(Y,X,Z)$ in population $s$, and let $p_s(y,z)$ denote the corresponding marginal density of $(Y,Z)$. The assumption allows the distribution of the observed variables $(Y,Z)$ to differ across populations, while requiring the conditional distribution of $X$ given $(Y,Z)$ to be invariant:
\begin{equation}\label{assum:joint pdf}
    p_s(y,x,z)=p_s(y,z)p(x\mid y,z),
    \qquad s\in\{0,1\}.
\end{equation}
Equivalently, if $S$ denotes the population indicator, this condition can be written as $X \perp S \mid (Y,Z)$, and therefore has a missing-at-random (MAR)-type interpretation. In the present two-sample setting, however, \(X\) is unobserved for every individual in the target population, so \(p(x\mid y,z)\) must be learned from the source data.

Let \(g\) be a known inverse link function and let $\beta:=(\beta_x^\t,\beta_z^\t)^\t\in\R^{p+q}$. We define the target parameter \(\beta_0\) as the unique solution to the population estimating equation
\begin{equation}\label{eq:estimating_eq}
    \Psi_0(\beta) := \E_0\{\varphi(Y,X,Z;\beta)\} = 0_{p+q}, 
\end{equation}
where
$
    \varphi(Y,X,Z;\beta) = (X^\t,Z^\t)^\t \{Y-g(X^\t\beta_x+Z^\t\beta_z)\}.
$ 
Here, \(0_d\) denotes the \(d\)-dimensional zero vector, and we omit the dimension subscript hereafter when clear from context. Different choices of \(g\) yield familiar generalized regression structures, including linear, logistic, and probit regression. If the target conditional mean is correctly specified as
\[
    \E_0(Y\mid X,Z) = g(X^\t\beta_{x,0}+Z^\t\beta_{z,0}), 
\]
then \(\beta_0\) coincides with the corresponding generalized linear model (GLM) coefficient. More generally, Eq~\eqref{eq:estimating_eq} defines a target-population working regression parameter when the conditional mean model is misspecified, provided that the population estimating equation admits a unique and regular solution.

If a complete i.i.d. sample of $(Y,X,Z)$ from the target population is available, $\beta_0$ could be estimated directly by replacing the expectation in Eq~\eqref{eq:estimating_eq} with its empirical counterpart. In the CMC setting, however, this estimating equation cannot be constructed because $X$ is unobserved for every individual in the target population. Although the source data contain fully observed $(Y,X,Z)$, simply replacing $\E_0$ by $\E_1$ generally does not recover the target parameter because the distribution of $(Y,Z)$ may differ across populations.

An additional challenge arises from the nonlinear estimating equation. Unlike linear regression, generalized regression involves parameter-indexed conditional functionals of the missing covariates, such as $\E\{g(X^\t\beta_x+Z^\t\beta_z)\mid Y,Z\}$ and $\E\{Xg(X^\t\beta_x+Z^\t\beta_z)\mid Y,Z\}$. Thus, imputing $X$ or a fixed collection of its low-order conditional moments is generally insufficient. This motivates the transfer-learning strategies developed below.

%%%%%%%%%%%%%%%%%%%%%%%%
\subsection{Two Preliminary Transfer-Learning Strategies}\label{subsec:two_solutions}
%%%%%%%%%%%%%%%%%%%%%%%%
There are two natural transfer-learning strategies for estimating the target parameter \(\beta_0\) under the CMC setting: one is based on importance weighting, and the other is based on the idea of the missing data imputation.

First, we discuss the solution based on importance weighting.
Let 
\begin{equation}\label{eq:ImportanceWeight}
    \w(y,z)=\frac{p_0(y,z)}{p_1(y,z)}
\end{equation}
be the density ratio of $(Y,Z)$ between the two populations. We assume that $p_0(y,z)$ is absolutely continuous with respect to $p_1(y,z)$ and that $\w(y,z)$ is uniformly bounded on the support of the target distribution. 
Under the sub-population shift assumption~\eqref{assum:joint pdf}, importance weighting gives the estimating equation:
\begin{equation}\label{eq:iw_estimating}
    \Psi_{\rm IW}(\beta):=\E_1\{\w(Y,Z)\varphi(Y,X,Z;\beta)\} = 0.  
\end{equation}
If $\w(y,z)$ is known, Eq~\eqref{eq:iw_estimating} would provide a feasible estimating equation for $\beta_0$ based on the source data. Because the density ratio $\w(y,z)$ is typically unknown in practice, we introduce a parametric working model
\begin{equation}
    \omega(y,z):=\omega(y,z;\eta),
\end{equation}
where the nuisance parameter $\eta$ can be estimated from the observed data (see Section~\ref{subsec:densityratio} for details). Let $\widehat\eta$ denote the estimated nuisance parameter and define
$\widehat\omega(y,z):=\omega(y,z;\widehat\eta)$. A feasible estimating equation for $\beta_0$ is obtained from Eq~\eqref{eq:iw_estimating} by replacing $\w(Y,Z)$ with $\widehat\omega(Y,Z)$ and the population expectation with its empirical counterpart. We refer to this method as the \emph{importance weighting} (IW) method, and denote the corresponding estimator by $\widehat\beta_{\rm IW}:=(\widehat\beta_{x,\rm IW}^\t,\widehat\beta_{z,\rm IW}^\t)^\t$.

Next, we discuss the solution based on missing data imputation. Define
\begin{equation*}
    \bar\varphi(Y,Z;\beta):=\E\{\varphi(Y,X,Z;\beta)\mid Y,Z\},
\end{equation*}
where we omit the subscripts of $\E$ because, under the sub-population shift assumption~\eqref{assum:joint pdf}, the conditional distribution $p(x\mid y,z)$ is shared by the two populations.
By the law of total expectation, Eq~\eqref{eq:estimating_eq} can be equivalently written as
\begin{equation}\label{eq:imp_estimating}
    \Psi_{\rm IMP}(\beta):=\E_0\{\bar\varphi(Y,Z;\beta)\} = 0.
\end{equation}
For the generalized linear estimating equation, define
\begin{align}\label{eq:mu_iota}
    \mu_1(y,z) &= \E(X \mid Y=y, Z=z)=\int x p(x\mid y,z)dx,\nonumber\\
    \mu_2(y,z;\beta) &= \E\{g(X^\t\beta_x+Z^\t\beta_z) \mid Y=y, Z=z\}=\int g(x^\t\beta_x+z^\t\beta_z) p(x\mid y,z)dx,\ \ \\
    \mu_3(y,z;\beta) &= \E\{Xg(X^\t\beta_x+Z^\t\beta_z) \mid Y=y, Z=z\}=\int xg(x^\t\beta_x+z^\t\beta_z) p(x\mid y,z)dx.\nonumber
\end{align}
Then,
\[
\bar\varphi(Y,Z;\beta)
=
\begin{pmatrix}
\mu_1(Y,Z)Y-\mu_3(Y,Z;\beta)\\
Z\{Y-\mu_2(Y,Z;\beta)\}
\end{pmatrix}.
\]
If $\{\mu_\iota\}_{1\leq\iota\leq3}$ are known, Eq~\eqref{eq:imp_estimating} would provide a feasible estimating equation for $\beta_0$ using the observed $(Y,Z)$ in the target population.

Because $\{\mu_\iota\}_{1\leq\iota\leq3}$ are unknown in practice, we estimate them through a parametric working model for $p(x\mid y,z)$:
\begin{equation}\label{eq:WorkingModel4CondDensity}
    q(x\mid y,z):= q(x\mid y,z;\gamma),
\end{equation}
where the nuisance parameter $\gamma$ is estimated from the source data (see Section~\ref{subsec:imputation} for details). The working model $q(x\mid y,z)$ induces the following working conditional functionals:
\begin{align}\label{eq:imputation}
    & m_1(y,z) = \int x q(x\mid y,z)dx,\qquad m_2(y,z;\beta) = \int g(x^\t\beta_x+z^\t\beta_z)q(x\mid y,z)dx,\nonumber\\
    & m_3(y,z;\beta) = \int xg(x^\t\beta_x+z^\t\beta_z)q(x\mid y,z)dx,
\end{align}
We omit the dependence of $\{m_\iota\}_{1\leq\iota\leq3}$ on $\gamma$ for notational simplicity.
If $q(x\mid y,z)$ is correctly specified for $p(x\mid y,z)$, then the induced working conditional functionals $\{m_\iota\}_{1\leq\iota\leq3}$ coincide with $\{\mu_\iota\}_{1\leq\iota\leq3}$. 
Replacing the unknown conditional functionals in Eq~\eqref{eq:imp_estimating} by their fitted working counterparts yields the \emph{imputation} (IMP) estimator $\widehat\beta_{\rm IMP}:=(\widehat\beta_{x,\rm IMP}^\t,\widehat\beta_{z,\rm IMP}^\t)^\t$. Its practical implementation is described in Section~\ref{subsec:imputation}.

%%%%%%%%%%%%%%%%%%%%%%%%
\subsection{Doubly Robust Transfer Learning}\label{subsec:DR_solution}
%%%%%%%%%%%%%%%%%%%%%%%%
The IW and IMP methods each rely on the correct specification of one nuisance component. To obtain an estimator that remains valid when either the density ratio model or the conditional covariate model is correctly specified, we combine importance weighting and conditional imputation through an augmented estimating equation.

Define the working conditional score induced by $q(x\mid y,z)$ as
\[
    \bar\varphi_q(y,z;\beta)
    :=
    \int \varphi(y,x,z;\beta)q(x\mid y,z)\,dx.
\]
By Eq~\eqref{eq:imputation}, this quantity can be written as
\[
    \bar\varphi_q(y,z;\beta)
    =
    \begin{pmatrix}
        m_1(y,z)y-m_3(y,z;\beta)\\
        z\{y-m_2(y,z;\beta)\}
    \end{pmatrix}.
\]
We then consider the augmented estimating equation
\begin{equation}\label{eq:DR_estimating}
    \Psi_{\rm DR}(\beta;\omega,q)
    :=
    \E_0\{\bar\varphi_q(Y,Z;\beta)\}
    +
    \E_1\left[
        \omega(Y,Z)
        \{\varphi(Y,X,Z;\beta)-\bar\varphi_q(Y,Z;\beta)\}
    \right]
    =0.
\end{equation}
The key property of Eq~\eqref{eq:DR_estimating} is its double robustness, formalized in the following theorem. Its proof is detailed in Appendix~1.1.

\begin{theorem}[Double robustness]\label{thm:DR}
Under the sub-population shift assumption~\eqref{assum:joint pdf}, if either
$
    \omega(y,z)=\w(y,z)
$
or 
$
    q(x\mid y,z)=p(x\mid y,z),
$
then
\[
    \Psi_{\rm DR}(\beta_0;\omega,q)=0.
\]
\end{theorem}

Theorem~\ref{thm:DR} shows that the augmented estimating equation remains unbiased at $\beta_0$ when either nuisance working model is correctly specified.
The corresponding sample estimating equation is obtained by replacing the expectations in Eq~\eqref{eq:DR_estimating} with their empirical counterparts and the nuisance functions with their estimated versions. We denote its solution by
$
    \widehat\beta_{\rm DR}
    :=
    (\widehat\beta_{x,\rm DR}^\t,
    \widehat\beta_{z,\rm DR}^\t)^\t,
$
and refer to it as the DRTL-CMC estimator. Its practical implementation is described in Section~\ref{sec:estimation}.

%%%%%%%%%%%%%%%%%%%%%%%%%%%%%%%%%%%%
\section{Practical Estimation}\label{sec:estimation}
%%%%%%%%%%%%%%%%%%%%%%%%%%%%%%%%%%%%
This section describes the practical implementation of the DRTL-CMC estimator. The procedure requires estimating two nuisance components: the density ratio \(\w(y,z)\) between the target and source distributions of \((Y,Z)\), and the conditional distribution \(p(x\mid y,z)\) of the target-missing covariates.

%%%%%%%%%%%%%%%%%%%%%%%%
\subsection{Specification and Estimation of Working Model $\omega(y,z)$}\label{subsec:densityratio}
%%%%%%%%%%%%%%%%%%%%%%%%
Following \citet{zhao2026augmented}, we specify the following working model for the density ratio $\w(y,z)$:
\begin{equation}\label{eq:densityratio}
    \omega(y,z;\eta) = \exp(y\eta_y+z^\t\eta_z),
\end{equation}
where $\eta=(\eta_y,\eta_z^\t)^\t$, with $\eta_y\in\R$ and $\eta_z\in\R^q$, denotes the nuisance parameter.

We define the population target of the working model~\eqref{eq:densityratio} as
\begin{align}\label{eq:barEta}
    \bar\eta
    =
    \arg\min_{\eta_y,\eta_z}
    \left[
        \E_1\{\exp(Y\eta_y+Z^\t\eta_z)\}
        -
        \E_0(Y\eta_y+Z^\t\eta_z)
    \right].
\end{align}
Its first-order condition is $\E_1\{\omega(Y,Z;\bar\eta)(Y,Z^\t)^\t\} - \E_0\{(Y,Z^\t)^\t\} =0.$
Under correct specification, $\omega(y,z;\bar\eta)=\w(y,z)$.

Replacing the population expectations in Eq~\eqref{eq:barEta} with their empirical counterparts yields 
\begin{align}
    \widehat\eta = \arg\min_{\eta_y\in\R,\eta_z\in\R^q} \l\{\frac{1}{n}\sum^n_{i=1} \exp\l(Y_i\eta_y+Z_i^\t\eta_z\r) -  \frac{1}{N}\sum^{n+N}_{i=n+1} \l(Y_i\eta_y+Z_i^\t\eta_z\r)\r\}.
\end{align}
Plugging $\widehat\eta$ into Eq~\eqref{eq:densityratio}, we obtain the fitted density ratio as
\begin{equation}\label{eq:omega_hat}
    \widehat\omega(y,z)= \omega(y,z;\widehat\eta) =\exp(y\widehat\eta_y+z^\t\widehat\eta_z).
\end{equation}

%%%%%%%%%%%%%%%%%%%%%%%%
\subsection{Specification and Estimation of Working Model $q(x\mid y,z)$}\label{subsec:imputation} 
%%%%%%%%%%%%%%%%%%%%%%%%
The specification of the working model $q(x\mid y,z)$ for $p(x\mid y,z)$ depends on the characteristics of the missing covariates $X$.
For continuous $X\in\R^p$, a convenient choice is a Gaussian working model,
\begin{equation}\label{eq:q_gaussian}
    q(\cdot\mid y,z;\gamma) =
 \cN_p\big(B^\t(y,z^\t)^\t,\Sigma\big),
\end{equation}
where $\gamma=(B,\Sigma)$ denotes the nuisance parameter.
For binary $X=(X_1,\cdots,X_p)^\t$, one may use a product Bernoulli working model
\begin{equation}\label{eq:q_binary}
q(x\mid y,z; \gamma)
=
\prod_{j=1}^{p}
\pi_j(y,z)^{x_j}
\{1-\pi_j(y,z)\}^{1-x_j},
\qquad
\operatorname{logit}\{\pi_j(y,z)\}
= (y,z^\t)\gamma_j,
\end{equation}
where $\gamma_j\in\R^{q+1}$ for $j=1,\ldots,p$ and $\gamma=(\gamma_1,\ldots,\gamma_p)\in\R^{(q+1)\times p}$
denotes the nuisance parameter matrix, and $\text{logit}(a)=\log\big(a/(1-a)\big)$ for given $a\in(0,1)$. This model can be fitted by separate logistic regressions in the source sample.

More generally, the conditional working distribution need not be specified through a standard parametric regression model. Motivated by the auxiliary density-ratio representation in \citet{yan2024transfer} and exponential-tilting density-ratio models such as those considered in \citet{zhao2026augmented}, one may instead specify a conditional exponential-tilting model
$$
    q(x\mid y,z;\gamma)
    =
    \frac{
        \exp\{h(x,y,z)^\t\gamma\}\widetilde p(x)
    }{
        \int
        \exp\{h(u,y,z)^\t\gamma\}\widetilde p(u)\,du
    },
$$
where $\widetilde p$ is a chosen reference density and $h(x,y,z)$ is a user-specified basis that allows the conditional
distribution of $X$ to vary with $(Y,Z)$. This provides a flexible alternative to the Gaussian and Bernoulli working models.

Regardless of the specific working model used, the nuisance parameter
$\gamma$ is estimated from the source data. Let $\widehat\gamma$ denote the resulting estimator. Under correct specification, $\widehat\gamma$ consistently estimates the corresponding conditional-model parameter. We then define
$
    \widehat q(x\mid y,z)
    =
    q(x\mid y,z;\widehat\gamma).
$
Given $\widehat q(x\mid y,z)$, the fitted conditional functionals can be evaluated analytically when closed-form expressions are available. Otherwise, they can be approximated by Monte Carlo integration. Let $\{X^{(k)}_{y,z}\}_{k=1}^K$
be i.i.d. draws from $\widehat q(x\mid y,z)$. We compute
\begin{align}\label{eq:imputation_hat}
    & \widehat m_1(y,z)
    =
    \frac{1}{K}\sum_{k=1}^K X^{(k)}_{y,z},
    \qquad
    \widehat m_2(y,z;\beta)
    =
    \frac{1}{K}\sum_{k=1}^K
    g\left(\beta_x^\t X^{(k)}_{y,z}+z^\t\beta_z\right),
    \notag \\
    & \widehat m_3(y,z;\beta)
    =
    \frac{1}{K}\sum_{k=1}^K
    X^{(k)}_{y,z}
    g\left(\beta_x^\t X^{(k)}_{y,z}+z^\t\beta_z\right).
\end{align}

%%%%%%%%%%%%%%%%%%%%%%%%
\subsection{Sample Estimating Equation for the DRTL-CMC Estimator}
%%%%%%%%%%%%%%%%%%%%%%%%
Given the fitted nuisance functions $\widehat\omega$ and $\widehat q$, define the fitted working conditional score
\[
    \bar\varphi_{\widehat q}(y,z;\beta)
    :=
    \int
    \varphi(y,x,z;\beta)
    \widehat q(x\mid y,z)\,dx.
\]
Equivalently, using the fitted conditional functionals
$\{\widehat m_\iota\}_{1\leq\iota\leq3}$,
\[
    \bar\varphi_{\widehat q}(y,z;\beta)
    =
    \begin{pmatrix}
        \widehat m_1(y,z)y-\widehat m_3(y,z;\beta)\\
        z\{y-\widehat m_2(y,z;\beta)\}
    \end{pmatrix}.
\]

The sample analogue of the augmented estimating equation~\eqref{eq:DR_estimating}, which defines the DRTL-CMC estimator, is
\begin{equation}\label{eq:DR_estimating_hat}
    \widehat\Psi_{\rm DR}(\beta)
    :=
    \frac{1}{N}
    \sum_{i=n+1}^{n+N}
    \bar\varphi_{\widehat q}(Y_i,Z_i;\beta)
    +
    \frac{1}{n}
    \sum_{i=1}^{n}
    \widehat\omega(Y_i,Z_i)
    \left\{
        \varphi(Y_i,X_i,Z_i;\beta)
        -
        \bar\varphi_{\widehat q}(Y_i,Z_i;\beta)
    \right\}
    =0.
\end{equation}
Thus, the DRTL-CMC estimator $\widehat\beta_{\rm DR}$ introduced in Section~\ref{sec:method} is the solution to Eq~\eqref{eq:DR_estimating_hat}.

%%%%%%%%%%%%%%%%%%%%%%%%%%%%%%%%%%%%
\section{Theoretical Properties}\label{sec:theory}
%%%%%%%%%%%%%%%%%%%%%%%%%%%%%%%%%%%%
In this section, we establish the consistency and asymptotic normality of the DRTL-CMC estimator and show that it attains the semiparametric efficiency bound when the working nuisance models are correctly specified. Throughout the theoretical analysis, we assume
that $p$ and $q$ are fixed and that, as $n,N\to\infty$, $n/(n+N)\to\rho\in(0,1).$

We first establish the consistency results for the DRTL-CMC estimator $\widehat\beta_{\rm DR}$ as follows: 
\begin{theorem}\label{thm:main-consistent}
Suppose that the sub-population shift assumption~\eqref{assum:joint pdf} holds and that at least one of the following two conditions is satisfied:
\begin{enumerate}
    \item (Correct density-ratio model.) $\w(y,z) = \omega(y,z;\eta_0)$ for some $\eta_0$;
    \item (Correct conditional-covariate model.) $p(x\mid y,z) = q(x\mid y,z;\gamma_0)$ for some $\gamma_0$.
\end{enumerate}
Then, under the regularity conditions stated in Appendix~1.2,
\begin{equation*}
    \|\widehat\beta_{\rm DR}-\beta_0\|_2 = o_p(1),
\end{equation*}
where $\|a\|_2$ denotes the $\ell_2$ norm of the vector $a$.
\end{theorem}

The following two corollaries respectively establish the consistency of $\widehat\beta_{\rm IW}$ under correct specification of $\omega(y,z)$ and the consistency of $\widehat\beta_{\rm IMP}$ under correct specification of $q(x \mid y,z)$.
\begin{corollary}\label{cor:iw-consistency}
Under the sub-population shift assumption~\eqref{assum:joint pdf}, the regularity conditions (see Appendix~1.2), and the correct specification of the density-ratio model (i.e., condition 1 of Theorem~\ref{thm:main-consistent}),
\begin{equation*}
    \|\widehat\beta_{\rm IW}-\beta_0\|_2 = o_p(1).
\end{equation*}
\end{corollary}

\begin{corollary}\label{cor:imp-consistency}
Under the sub-population shift assumption~\eqref{assum:joint pdf}, the regularity conditions (see Appendix~1.2), and the correct specification of the conditional-covariate model (i.e., condition 2 of Theorem~\ref{thm:main-consistent}), 
\begin{equation*}
    \|\widehat\beta_{\rm IMP}-\beta_0\|_2 = o_p(1).
\end{equation*}
\end{corollary}

We now present the main theoretical results concerning the asymptotic validity of our estimator $c^\t\widehat\beta_{\rm DR}$ for any vector $c \in \R^{p+q}$ in Theorem~\ref{thm:main}. Without loss of generality, we assume that $\|c\|_2 = 1$.
\begin{theorem}\label{thm:main}
    Under the conditions of Theorem~\ref{thm:main-consistent} and the additional regularity conditions for asymptotic linearity stated in Appendix~1.2, it holds that 
    \begin{align*}
        \sqrt{n} c^\t(\widehat\beta_{\rm DR} - \beta_0)
        = \frac{1}{\sqrt{n}}\sum^n_{i=1} F^1_i + \frac{\sqrt{n}}{N}\sum^{n+N}_{i=n+1} F^0_i + \sqrt{n}\xi_{\eta}^\t(\widehat\eta-\bar\eta) + \sqrt{n}\xi_{\gamma}^\t(\widehat\gamma-\bar\gamma) + o_p(1),
    \end{align*}
    where $F^1_i$ and $F^0_i$ are source- and target-sample influence components, respectively, and $\xi_\eta$ and $\xi_\gamma$ are expectations involving the limiting nuisance parameters, with their concrete forms provided in Appendix~1.7.
    As a result,  $\sqrt{n}\  c^\t (\widehat\beta_{\rm DR} - \beta_0)$ weakly converges to a mean-zero Gaussian distribution with finite variance. 
\end{theorem}

The first two terms in Theorem~\ref{thm:main} represent the source- and target-sample contributions, respectively. When the density-ratio model is correctly specified, $\xi_\gamma=0$, so estimation of $\gamma$ does not contribute to the first-order asymptotic expansion. Similarly, when the conditional-covariate model is correctly specified, $\xi_\eta=0$, so estimation of $\eta$ does not contribute at first order.

\begin{theorem}\label{thm:semiparametric_efficient}
When both nuisance models are correctly specified, $\widehat\beta_{\rm DR}$ is a semiparametrically efficient estimator for $\beta_0$ under the sub-population shift assumption~\eqref{assum:joint pdf}. The efficiency bound for $\sqrt{n+N}(\widehat\beta_{\rm DR}-\beta_0)$ is $J_0 V_0J_0^\top$, 
where \(J_0:=-\left\{\left.\partial\Psi_0(\beta)/\partial\beta^\t\right|_{\beta=\beta_0}\right\}^{-1}\), with the detailed form provided in Appendix~1.8, and
\[
    V_0
    =
    \frac{1}{1-\rho}
    \E_0\{\psi(Y,Z;\beta_0)^{\otimes 2}\}
    +
    \frac{1}{\rho}
    \E_1\!\left[
        \w^2(Y,Z)
        \{\phi(X,Y,Z;\beta_0)-\psi(Y,Z;\beta_0)\}^{\otimes 2}
    \right].
\]
Here, $\phi(x,y,z;\beta) = (x^\t,z^\t)^\t \{y - g(x^\t\beta_x+ z^\t\beta_z)\}$, $\psi(y,z;\beta) = \E\{\phi(X,y,z;\beta)\mid y,z\}$, and for any vector $a$, we write $\E(aa^\t)$ as $\E(a^{\otimes 2})$ for simplicity. 
\end{theorem}
Theorem~\ref{thm:semiparametric_efficient} establishes the semiparametric efficiency bound for the moment-defined target parameter \(\beta_0\) under the CMC setting and the sub-population shift assumption. The proofs of Theorems~\ref{thm:main-consistent}–\ref{thm:semiparametric_efficient} are provided in Appendices~1.4-1.8.

\begin{corollary}\label{cor1}
Under the assumptions of Theorems~\ref{thm:main}–\ref{thm:semiparametric_efficient}, it holds that
\[
    \sqrt{n+N}\, 
    (\widehat\beta_{\rm DR} - \beta_0)
    \xrightarrow{d} N(0,\, J_0 V_0 J_0^\t).
\]
Equivalently,
$
    \sqrt{n}\,
    (\widehat\beta_{\rm DR} - \beta_0)
    \xrightarrow{d}
    N\big(0,\, \rho J_0 V_0 J_0^\top\big).
$
\end{corollary}

%%%%%%%%%%%%%%%%%%%%%%%%%%%%%%%%%%%%
\section{Simulation Studies}\label{sec:simulation}
%%%%%%%%%%%%%%%%%%%%%%%%%%%%%%%%%%%%
%%%%%%%%%%%%%%%%%%%%%%%%
\subsection{Simulation Settings}
In this section, we evaluate the finite sample performance of the DRTL-CMC estimator. We set the total sample size $n+N\in\{1500,3000\}$ and $(p,q)=(1,3)$, so that $X\in\R$ and $Z=(1,Z_2,Z_3)^\t$. We generate $(Z_{2i},Z_{3i})^\t\sim N_2(0,\Sigma_z),\Sigma_z=(0.2^{|j-k|})\in\R^{2\times 2}$ and set $Z_i=(1,Z_{2i},Z_{3i})^\t$. Conditional on $Z_i$, we generate the binary outcome $Y_i$ based on $\text{logit}\{P(Y_i=1\mid Z_i)\} = Z^\t_i(0.5,0.8,-0.3)$. 
We consider two mechanisms for generating $X_i$:
    \begin{align*}
        \mathrm{M}_{\rm cor}: \ & X_i = 1 + 0.4Y_i - 0.6Z_{2i} + 0.2Z_{3i} + \epsilon_i,\\
        \mathrm{M}_{\rm mis}: \ & X_i = 1 + 0.4Y_i - 0.6Z_{2i} + 0.2Z_{3i} + Z_{2i}^2 + \epsilon_i,
    \end{align*}
where $\epsilon_i\sim N(0,0.5^2)$ independently for $i=1,\ldots,n+N$. In both cases, we use the Gaussian working model in Eq~\eqref{eq:q_gaussian} for $p(x\mid y,z)$. This model is correctly specified under $\mathrm{M}_{\rm cor}$ but misspecified under $\mathrm{M}_{\rm mis}$, because the latter introduces the quadratic term $Z_2^2$ into the conditional mean of $X$ given $(Y,Z)$, which is omitted from the working model.

After generating $n+N$ complete observations of $(Y,X,Z)$, we generate a population membership indicator $S_i$, where $S_i=1$ denotes the source population and $S_i=0$ denotes the target population. We consider two membership mechanisms:
    \begin{align*}
        \mathrm{W}_{\rm cor}: \ & \text{logit}\{P(S_i=1\mid Y_i,Z_i)\} =  - 0.8 + 0.8Y_i - 0.2Z_{2i} + 0.3Z_{3i},\\
        \mathrm{W}_{\rm mis}: \ & \text{logit}\{P(S_i=1\mid Y_i,Z_i)\} = - 0.8 + 0.8Y_i - 0.2Z_{2i} + 0.3Z_{3i} - 0.3Y_iZ_{3i}.
    \end{align*}
Let $n$ and $N$ denote the resulting source and target sample sizes, respectively. For notational convenience, we reorder the observations so that the source observations are indexed by $1,\ldots,n$ and the target observations by $n+1,\ldots,n+N$. We then remove $X_i$ from all target observations to create the CMC setting.
In both membership mechanisms, we choose the working density-ratio model $\omega(y,z;\eta)$ in Eq~\eqref{eq:densityratio} for $\w(y,z)$. By Bayes' rule, 
$\w(y,z) = P(S=1)P(S=0\mid Y=y,Z=z)/\{P(S=0)P(S=1\mid Y=y,Z=z)\}$.
It follows that $\omega(y,z;\eta)$ is correctly specified under $\mathrm{W}_{\rm cor}$ but misspecified under $\mathrm{W}_{\rm mis}$, because the latter induces a $YZ_3$ interaction in the log density ratio that is omitted from Eq~\eqref{eq:densityratio}.

Combining the conditional-covariate and density-ratio mechanisms, we consider three configurations:
(I) $\mathrm{M}_{\rm cor}$ and $\mathrm{W}_{\rm cor}$, (II) $\mathrm{M}_{\rm mis}$ and $\mathrm{W}_{\rm cor}$, and (III) $\mathrm{M}_{\rm cor}$ and $\mathrm{W}_{\rm mis}$. Because $S_i$ is generated conditionally on $(Y_i,Z_i)$ only, $X\perp S\mid(Y,Z)$ holds by construction. Thus, all three configurations satisfy the sub-population shift assumption while allowing different patterns of nuisance-model specification. The simulation design follows a similar spirit to \citet{cai2025semi,zhao2026augmented}.

Our inferential target is the population parameter \(\beta_0\) defined by Eq~\eqref{eq:estimating_eq}. For each configuration, we generate an independent target-population sample of size \(10^6\) with $(Y,X,Z)$ fully observed and fit the corresponding working logistic regression. The resulting coefficient vector provides a high-precision Monte Carlo approximation to $\beta_0$ and serves as the benchmark for computing bias and root mean squared error (RMSE).

For reference, we also include two procedures developed for conventional missing-covariate settings: inverse probability weighting (IPW) \citep{lipsitz1999weighted} and the predict-then-debias (PTD) method \citep{kluger2025prediction}. These approaches require the missing covariate to be observed for at least a subset of the analysis population and are therefore not directly designed for the CMC setting. To implement them here, we pool the observed source and target data and ignore source--target heterogeneity.

%%%%%%%%%%%%%%%%%%%%%%%%
\subsection{Results}
For each setting, we conduct 200 Monte Carlo replications. Within each replication, standard errors are estimated using 500 bootstrap resamples. For the given estimator $\widehat\beta$, we report the empirical average bias, RMSE, average bootstrap standard error, and empirical coverage probability of the nominal 95\% confidence interval. We present the statistical inference results for $\beta$ under $n+N=1500$ in Tables~\ref{tab:point}-\ref{tab:var}, and the results for $\beta$ under $n+N=3000$ in Tables~A1-A2 in Appendix~2.

\begin{table}[!ht]
  \centering
  \caption{Point estimator results for the target parameter $\beta_0$ with total sample size $n+N=1500$.}\label{tab:point}
  \sisetup{
    table-format=-1.2,
    table-align-text-pre=false,
    table-align-text-post=false,
  }
  \resizebox{\columnwidth}{!}{
  \begin{threeparttable}
  \begin{tabular}{
    lrrrrr
    S[table-format=1.3]
    S[table-format=1.3]
    S[table-format=1.3]
    S[table-format=1.3]
    S[table-format=1.3]
  }
    \toprule
    & \multicolumn{5}{c}{\textbf{Average Bias}} & \multicolumn{5}{c}{\textbf{RMSE}} \\
    \cmidrule(lr){2-6} \cmidrule(lr){7-11}
    \textbf{$\beta_0$} & {\textbf{IPW}} & {\textbf{PTD}} & {\textbf{IW}} & {\textbf{IMP}} & {\textbf{DR}} & {\textbf{IPW}} & {\textbf{PTD}} & {\textbf{IW}} & {\textbf{IMP}} & {\textbf{DR}} \\
    \midrule
    \multicolumn{11}{c}{Configuration (I): $\mathrm{M}_{\rm cor}$ and $\mathrm{W}_{\rm cor}$} \\[2pt]
    $\beta_x = 1.603$      & 0.010 & 4.120 & 0.004 & 0.000 & -0.106 & 0.172 & 8.038 & 0.177 & 0.166 & 0.432 \\ 
    $\beta_{z_1} = -0.948$ & -1.280 & -5.193 & -0.013 & -0.005 & 0.120 & 1.303 & 9.977 & 0.226 & 0.220 & 0.529 \\ 
    $\beta_{z_2} = 1.728$  & 0.038 & 1.176 & -0.001 & 0.012 & -0.054 & 0.147 & 2.902 & 0.150 & 0.143 & 0.259 \\ 
    $\beta_{z_3} = -0.569$ & -0.051 & -0.208 & 0.005 & -0.008 & 0.014 & 0.097 & 0.719 & 0.090 & 0.101 & 0.111 \\ 
    \midrule
    \multicolumn{11}{c}{Configuration (II): $\mathrm{M}_{\rm mis}$ and $\mathrm{W}_{\rm cor}$} \\[2pt]
    $\beta_x = 0.281$      & -0.036 & 0.031 & 0.004 & -0.102 & 0.029 & 0.075 & 0.201 & 0.080 & 0.114 & 0.201 \\ 
    $\beta_{z_1} = 0.415$  & -1.184 & -0.068 & -0.011 & 0.167 & -0.048 & 1.199 & 0.429 & 0.170 & 0.216 & 0.371 \\ 
    $\beta_{z_2} = 1.037$  & 0.011 & -0.002 & 0.003 & -0.115 & 0.044 & 0.136 & 0.164 & 0.153 & 0.171 & 0.225 \\ 
    $\beta_{z_3} = -0.304$ & -0.046 & -0.006 & 0.003 & 0.009 & -0.018 & 0.086 & 0.096 & 0.082 & 0.093 & 0.095 \\ 
    \midrule
    \multicolumn{11}{c}{Configuration (III): $\mathrm{M}_{\rm cor}$ and $\mathrm{W}_{\rm mis}$} \\[2pt]
    $\beta_x = 1.602$      & 0.011 & 4.383 & 0.010 & 0.002 & -0.089 & 0.173 & 7.513 & 0.175 & 0.170 & 0.412 \\ 
    $\beta_{z_1} = -0.940$ & -1.279 & -5.484 & -0.019 & -0.005 & 0.102 & 1.301 & 9.291 & 0.238 & 0.232 & 0.507 \\ 
    $\beta_{z_2} = 1.729$  & 0.040 & 1.332 & 0.006 & 0.008 & -0.046 & 0.145 & 2.729 & 0.140 & 0.155 & 0.254 \\ 
    $\beta_{z_3} = -0.424$ & -0.307 & -0.182 & -0.290 & 0.001 & 0.020 & 0.318 & 0.607 & 0.302 & 0.098 & 0.116 \\ 
    \bottomrule
  \end{tabular}
\begin{tablenotes}
\footnotesize
\item[] IPW: inverse probability weighting; PTD: predict-then-debias; IW: importance weighting; IMP: imputation; DR: doubly robust (DRTL-CMC). $\mathrm{M}_{\rm cor}/\mathrm{M}_{\rm mis}$ indicate correct/misspecified conditional-covariate models; $\mathrm{W}_{\rm cor}/\mathrm{W}_{\rm mis}$ indicate correct/misspecified density-ratio models.
\end{tablenotes}
\end{threeparttable}
}
\end{table}

\begin{table}[!ht]
  \centering
  \caption{Variance estimator results for the target parameter $\beta_0$ with total sample size $n+N=1500$.}\label{tab:var}
  \sisetup{
    table-format=-1.2,
    table-align-text-pre=false,
    table-align-text-post=false,
  }
  \resizebox{\columnwidth}{!}{
  \begin{threeparttable}
  \begin{tabular}{
    lrrrrr
    S[table-format=1.3]
    S[table-format=1.3]
    S[table-format=1.3]
    S[table-format=1.3]
    S[table-format=1.3]
  }
    \toprule
    & \multicolumn{5}{c}{\textbf{Standard Error}} & \multicolumn{5}{c}{\textbf{Coverage Probability}} \\
    \cmidrule(lr){2-6} \cmidrule(lr){7-11}
    \textbf{$\beta_0$} & {\textbf{IPW}} & {\textbf{PTD}} & {\textbf{IW}} & {\textbf{IMP}} & {\textbf{DR}} & {\textbf{IPW}} & {\textbf{PTD}} & {\textbf{IW}} & {\textbf{IMP}} & {\textbf{DR}} \\
    \midrule
    \multicolumn{11}{c}{Configuration (I): $\mathrm{M}_{\rm cor}$ and $\mathrm{W}_{\rm cor}$} \\[2pt]
    $\beta_x = 1.603$      & 0.173 & 14.847 & 0.177 & 0.094 & 0.345 & 0.940 & 0.850 & 0.955 & 0.705 & 0.910 \\ 
    $\beta_{z_1} = -0.948$ & 0.246 & 18.397 & 0.232 & 0.151 & 0.441 & 0.005 & 0.850 & 0.950 & 0.845 & 0.910 \\ 
    $\beta_{z_2} = 1.728$  & 0.146 & 5.721 & 0.150 & 0.110 & 0.243 & 0.945 & 0.860 & 0.950 & 0.870 & 0.940 \\ 
    $\beta_{z_3} = -0.569$ & 0.092 & 1.379 & 0.096 & 0.100 & 0.114 & 0.940 & 0.895 & 0.980 & 0.955 & 0.960 \\ 
    \midrule
    \multicolumn{11}{c}{Configuration (II): $\mathrm{M}_{\rm mis}$ and $\mathrm{W}_{\rm cor}$} \\[2pt]
    $\beta_x = 0.281$      & 0.064 & 0.217 & 0.080 & 0.026 & 0.126 & 0.910 & 0.975 & 0.955 & 0.180 & 0.855 \\ 
    $\beta_{z_1} = 0.415$  & 0.176 & 0.458 & 0.175 & 0.112 & 0.278 & 0.000 & 0.970 & 0.970 & 0.630 & 0.900 \\ 
    $\beta_{z_2} = 1.037$  & 0.129 & 0.170 & 0.146 & 0.109 & 0.143 & 0.945 & 0.950 & 0.925 & 0.735 & 0.840 \\ 
    $\beta_{z_3} = -0.304$ & 0.080 & 0.105 & 0.085 & 0.100 & 0.100 & 0.940 & 0.960 & 0.975 & 0.970 & 0.955 \\ 
    \midrule
    \multicolumn{11}{c}{Configuration (III): $\mathrm{M}_{\rm cor}$ and $\mathrm{W}_{\rm mis}$} \\[2pt]
   $\beta_x = 1.602$      & 0.170 & 14.420 & 0.172 & 0.094 & 0.335 & 0.935 & 0.890 & 0.960 & 0.705 & 0.930 \\ 
   $\beta_{z_1} = -0.940$ & 0.243 & 17.518 & 0.227 & 0.152 & 0.428 & 0.000 & 0.885 & 0.950 & 0.815 & 0.920 \\ 
   $\beta_{z_2} = 1.729$  & 0.144 & 5.348 & 0.146 & 0.109 & 0.241 & 0.940 & 0.905 & 0.965 & 0.810 & 0.940 \\ 
   $\beta_{z_3} = -0.424$ & 0.092 & 1.700 & 0.093 & 0.098 & 0.119 & 0.050 & 0.935 & 0.085 & 0.940 & 0.965 \\ 
    \bottomrule
  \end{tabular}
\begin{tablenotes}
\footnotesize
\item[] IPW: inverse probability weighting; PTD: predict-then-debias; IW: importance weighting; IMP: imputation; DR: doubly robust (DRTL-CMC). $\mathrm{M}_{\rm cor}/\mathrm{M}_{\rm mis}$ indicate correct/misspecified conditional-covariate models; $\mathrm{W}_{\rm cor}/\mathrm{W}_{\rm mis}$ indicate correct/misspecified density-ratio models.
\end{tablenotes}
\end{threeparttable}
}
\end{table}

As seen from Tables~\ref{tab:point}-\ref{tab:var}, the IPW method performs poorly for $\beta_{z_1}$ across all three configurations and has a large bias for $\beta_{z_3}$ in Configuration (III). The PTD method shows substantial bias for several coefficients in Configurations (I) and (III),  together with large standard errors. As expected, the reference IPW and PTD procedures can exhibit substantial bias or variability when applied by pooling the two heterogeneous populations, reflecting their mismatch with the CMC data structure.

When both nuisance models are correctly specified (Configuration (I)), IW and IMP have the smallest RMSEs, whereas DRTL-CMC exhibits greater finite-sample variability and somewhat larger bias for some coefficients. The corresponding results for $n+N=3000$ show improved finite-sample performance, as reported in Tables~A1-A2 in Appendix~2. When $q(x\mid y,z)$ is misspecified (Configuration (II)), IMP often exhibits larger bias and RMSE than IW and DRTL-CMC. When $\omega(y,z)$ is misspecified (Configuration (III)), IW generally shows larger bias and RMSE than IMP and DRTL-CMC. Overall, DRTL-CMC substantially reduces the bias caused by single nuisance-model misspecification. These patterns are consistent with the double robustness property in Theorem~\ref{thm:DR}. For the variance estimator, DRTL-CMC generally has larger standard errors than IW and IMP, which reflects the additional variability introduced by the augmentation term. Regarding coverage, IMP has coverage rates below the nominal 95\% level in several cases, while IW shows severe undercoverage under density-ratio misspecification in Configuration (III) (for $\beta_{z_3}$, the coverage probability 8.5\%). In contrast, DRTL-CMC generally avoids such severe coverage deterioration under single nuisance-model misspecification, although its finite-sample coverage remains below the nominal level for some coefficients.

%%%%%%%%%%%%%%%%%%%%%%%%%%%%%%%%%%%%
\section{Validation Studies using UK Biobank Data}\label{sec:data analysis}
%%%%%%%%%%%%%%%%%%%%%%%%%%%%%%%%%%%%
\subsection{Data and Validation Design}
We evaluate the proposed method using UK Biobank data
(\href{https://www.ukbiobank.ac.uk/}{www.ukbiobank.ac.uk}). To construct a validation setting with completely missing covariates, we partition the data into source and target populations and artificially mask a key covariate in the target population. Because this covariate is observed before masking, the complete target data provide a benchmark for the target parameter $\beta_0$ defined by Eq~\eqref{eq:estimating_eq}. We compare DRTL-CMC with the two preliminary methods, IW and IMP.

Our response of interest $Y$ is obesity, defined as one when body mass index (BMI) is at least 30 and zero otherwise. We analyze 7,919 White British participants (3,859 males and 4,060 females) under two artificially constructed missingness scenarios. In \textbf{Case I}, the target population consists of 2,991 individuals aged 65 or older, while the source population consists of 4,928 individuals younger than 65. The observed covariates $Z$ are total energy and sex. In \textbf{Case II}, the target population consists of 3,150 previous or current smokers, while the source population consists of 4,759 never-smokers. Here, $Z$ includes total energy, sex, and age, consistent with BMI-related epidemiological studies \citep{lovejoy2009sex,arem2013healthy}. In both cases, the polygenic risk score for BMI serves as $X$, which may be unavailable in some studies or cohorts. We standardize total energy and center age and $X$ to have mean zero in the subsequent analyses.

Because $X$ is observed in this artificial validation design, we conduct exploratory diagnostics related to the sub-population shift assumption~\eqref{assum:joint pdf}. We first regress $X$ on $(Y,Z)$ in the pooled sample and compare the resulting residual distributions between the source and target populations. Kolmogorov--Smirnov tests show no statistically significant differences in either case (Case I: $p$-value $=0.781$; Case II: $p$-value $=0.567$). We also fit separate regressions of $X$ on $(Y,Z)$ in the two populations and obtain broadly similar coefficient estimates. Detailed results are provided in Table~A3 in Appendix~3. These diagnostics assess selected aspects of source--target comparability, but they do not verify equality of the full conditional distribution $p(x\mid y,z)$ required by the identifying assumption.

%%%%%%%%%%%%%%%%%%%%%%%%
\subsection{Complete-Data Benchmark Analysis}
%%%%%%%%%%%%%%%%%%%%%%%%
We begin by fitting the working logistic regression separately in the target and source populations using the fully observed \(X\) (Table~\ref{tab:summary}). The complete-data target fit is the empirical analogue of Eq~\eqref{eq:estimating_eq} and therefore provides a benchmark for the target parameter \(\beta_0\), while the source fit provides a reference for assessing source--target differences. Comprehensive results are provided in Table~A4 in Appendix~3.

\begin{table}[!ht]
  \centering
  \caption{Complete-data benchmark results with $X$ observed in both populations.}\label{tab:summary}
  \sisetup{
    table-format=-1.3,
    table-align-text-pre=false,
    table-align-text-post=false,
  }
  \begin{threeparttable}
  \begin{tabular}{
    l
    l
    S[table-format=-1.3]
    r
    l
    S[table-format=-1.3]
    r
  }
    \toprule
    & \textbf{Covariate} & {\textbf{Estimate}} & {\textbf{$p$-value}} & \textbf{Covariate} & {\textbf{Estimate}} & {\textbf{$p$-value}} \\
    \midrule
    \multirow{7}{*}{Case I} & \multicolumn{3}{c}{\textbf{Target (age $\geq$ 65)}} & \multicolumn{3}{c}{\textbf{Source (age $<$ 65)}} \\
    \cmidrule(lr){2-4} \cmidrule(l){5-7}
    & $X$       & 0.486 & $<0.001$  & $X$ & 0.538 & $<0.001$ \\
    & intercept & -1.507 & $<0.001$ & intercept & -1.595 & $<0.001$ \\
    & energy    & 0.050 & 0.307    & energy & 0.059 & 0.115 \\
    & sex       & 0.133 & 0.172    & sex & 0.221 & 0.004 \\
    \midrule
    \multirow{8}{*}{Case II} & \multicolumn{3}{c}{\textbf{Target (smoker)}} & \multicolumn{3}{c}{\textbf{Source (non-smoker)}} \\
    \cmidrule(lr){2-4} \cmidrule(l){5-7}
    & $X$       & 0.487 & $<0.001$  & $X$ & 0.544 & $<0.001$ \\
    & intercept & -1.376 & $<0.001$ & intercept & -1.702 & $<0.001$ \\
    & energy    & 0.006 & 0.893    & energy & 0.098 & 0.014 \\
    & sex       & 0.337 & $<0.001$  & sex & 0.038 & 0.641 \\
    & age       & -0.006 & 0.331   & age & 0.008 & 0.148 \\
    \bottomrule
  \end{tabular}
\begin{tablenotes}[flushleft]
\footnotesize
\item[] $X$ denotes the polygenic risk score for BMI.
\end{tablenotes}
  \end{threeparttable}
\end{table}

As shown in Table~\ref{tab:summary}, the estimated regression coefficients differ between the source and target populations in both cases. 
In Case I, the coefficient of $X$ is smaller in the target population (age $\geq$ 65) than in the source population (age $<$ 65), suggesting a weaker association between \(X\) and obesity among older individuals. The estimated coefficients for energy and sex also differ between the two populations. In Case II, the coefficient of $X$ for the target population (smoker) is also smaller than that of the source population (non-smoker), suggesting that the association between the polygenic risk score for BMI and obesity may be weaker among smokers. Additionally, the estimated coefficients for energy and sex differ between the two populations, and the coefficients for age have opposite signs.

These differences illustrate the heterogeneity between the source and target populations and reveal the limitations of directly applying analyses from the source data to the target population. Such disparities emphasize the need for tailored methods that account for population-specific variations.

%%%%%%%%%%%%%%%%%%%%%%%%
\subsection{Comparison of Estimation and Inference Across Methods}
Given that our simulations show the IPW and PTD methods are misaligned with the CMC setting and perform poorly under population heterogeneity, we compare the DRTL-CMC method only with the preliminary IW and IMP methods when $X$ is completely missing in the target data. 
We consider the linear imputation model $X \sim Y + energy + sex$ for Case I and $X \sim Y + energy + sex + age$ for Case II. Table~\ref{tab:dataresults} summarizes the differences between the point estimates and the complete-data target benchmark, bootstrap standard errors ($B=500$), $p$-values for testing the null hypothesis that the covariate coefficients equal zero, and nominal 95\% confidence intervals.

\begin{table}[!ht]
  \centering
  \caption{Comparison of estimation and inference for the target population with $X$ masked.}\label{tab:dataresults}
  \sisetup{
    table-format=-1.3,
    table-align-text-pre=false,
    table-align-text-post=false,
    input-open-uncertainty=(,
    input-close-uncertainty=),
    table-space-text-pre=(,
    table-space-text-post=),
  }
  \begin{threeparttable}
  \begin{tabular}{
    @{}
    c
    c
    l
    c 
    S[table-format=1.3]
    c
    r
    @{}
  }
    \toprule
    & \multicolumn{1}{c}{\textbf{Method}} & \multicolumn{1}{c}{\textbf{Covariate}} & {\textbf{Difference}} & {\textbf{SE}} & \multicolumn{1}{c}{\textbf{95\% CI}} & {\textbf{$p$-value}} \\
    \midrule
    \multirow{13}{*}{Case I} & \multirow{4}{*}{IW} 
    & $X$         & 0.052 & 0.040 & (0.461, 0.616) & $<0.001$\\
    & & intercept & -0.023 & 0.049 & (-1.625, -1.435) & $<0.001$\\
    & & energy    & 0.008 & 0.039 & (-0.019, 0.136) & 0.139 \\
    & & sex       & 0.087 & 0.080 & (0.064, 0.377) & 0.006 \\
    \cmidrule{2-7}
    & \multirow{4}{*}{IMP} 
    & $X$         & 0.050 & 0.039 & (0.459, 0.613) & $<0.001$\\
    & & intercept & -0.019 & 0.048 & (-1.620, -1.431) & $<0.001$\\
    & & energy    & -0.012 & 0.051 & (-0.063, 0.139) & 0.457 \\
    & & sex       & 0.014 & 0.092 & (-0.033, 0.328) & 0.109 \\
    \cmidrule{2-7}
    & \multirow{4}{*}{DRTL-CMC} 
    & $X$         & 0.046 & 0.039 & (0.456, 0.608) & $<0.001$\\
    & & intercept & -0.018 & 0.049 & (-1.622, -1.428) & $<0.001$\\
    & & energy    & -0.012 & 0.052 & (-0.063, 0.139) & 0.458 \\
    & & sex       & 0.010 & 0.092 & (-0.036, 0.324) & 0.118 \\
    \midrule
    \multirow{16}{*}{Case II} & \multirow{5}{*}{IW} 
    & $X$         & 0.057 & 0.039 & (0.467, 0.621) & $<0.001$\\
    & & intercept & -0.023 & 0.048 & (-1.492, -1.306) & $<0.001$\\
    & & energy    & 0.090 & 0.040 & (0.016, 0.175) & 0.018 \\
    & & sex       & -0.299 & 0.083 & (-0.125, 0.201) & 0.649 \\
    & & age       & 0.014 & 0.006 & (-0.003, 0.019) & 0.175 \\
    \cmidrule{2-7}
    & \multirow{5}{*}{IMP} 
    & $X$         & 0.056 & 0.039 & (0.466, 0.620) & $<0.001$\\
    & & intercept & -0.027 & 0.048 & (-1.496, -1.309) & $<0.001$\\
    & & energy    & 0.001 & 0.044 & (-0.078, 0.093) & 0.865 \\
    & & sex       & -0.024 & 0.091 & (0.134, 0.492) & 0.001 \\
    & & age       & 0.000 & 0.006 & (-0.018, 0.007) & 0.358 \\
    \cmidrule{2-7}
    & \multirow{5}{*}{DRTL-CMC} 
    & $X$         & 0.029 & 0.037 & (0.443, 0.589) & $<0.001$\\
    & & intercept & -0.019 & 0.048 & (-1.490, -1.300) & $<0.001$\\
    & & energy    & -0.001 & 0.043 & (-0.079, 0.090) & 0.901 \\
    & & sex       & -0.029 & 0.091 & (0.130, 0.487) & 0.001 \\
    & & age       & 0.000 & 0.006 & (-0.018, 0.006) & 0.347 \\
    \bottomrule
  \end{tabular}
\begin{tablenotes}[flushleft]
\footnotesize
\item[] ``Difference'' denotes the point estimate minus the complete-data target benchmark; SE denotes the bootstrap standard error. 
\end{tablenotes}
  \end{threeparttable}
\end{table}

As shown in Table~\ref{tab:dataresults}, the DRTL-CMC estimates are close to the complete-data target benchmarks in both cases. The preliminary IMP method also yields similar point estimates, suggesting that the linear imputation models provide reasonable approximations in these artificial validation settings. In contrast, the preliminary IW estimator shows larger deviations from the complete-data benchmarks for several covariates in both cases. For sex, IW yields a significant association in Case I that is not observed in the benchmark analysis, while in Case II it fails to detect the significant association observed in the benchmark analysis. Overall, DRTL-CMC and IMP yield similar estimation and inference results in these validation settings, while the double robustness of DRTL-CMC offers additional protection against misspecification of either nuisance model, as illustrated in the simulation results (Table~\ref{tab:var}).

%%%%%%%%%%%%%%%%%%%%%%%%%%%%%%%%%%%%
\section{Application to NHANES Data}\label{sec:data analysis-nhanes} 
%%%%%%%%%%%%%%%%%%%%%%%%%%%%%%%%%%%%
In this section, we apply the DRTL-CMC method to a real-data example based on the National Health and Nutrition Examination Survey (NHANES) dataset (\href{https://wwwn.cdc.gov/nchs/nhanes/}{wwwn.cdc.gov/nchs/nhanes}), which is a nationally representative program in the United States that collects detailed information on the health, nutrition, and lifestyle of adults and children.

%%%%%%%%%%%%%%%%%%%%%%%%
\subsection{Data Introduction}
The target population consists of 3,624 Non-Hispanic White participants (1,752 females and 1,872 males) from the 2001–2002 NHANES dataset, while the source population consists of 1,827 Non-Hispanic White participants (913 females and 914 males) from the 2015–2016 NHANES dataset. Our objective is to investigate the association between obesity and sleep duration in the target population. The binary outcome $Y$ indicates obesity ($Y=1$ if BMI $\geq 30$ and $Y=0$ otherwise). The covariate of interest $X$ is average nightly sleep duration on weekdays or workdays, ranging from 2 to 13.5 hours. This variable is unavailable in the target 2001--2002 dataset but observed in the source 2015--2016 dataset. The covariate vector $Z$, observed in both populations, consists of total energy intake, sex, and age.

The target of inference is the working logistic regression parameter $\beta_0$ defined by Eq~\eqref{eq:estimating_eq}. We compare DRTL-CMC with the two preliminary methods, IW and IMP. Because $X$ is unobserved in the target sample, the sub-population shift assumption~\eqref{assum:joint pdf} cannot be empirically verified in this application. The results should therefore be interpreted conditional on the transportability of $X\mid(Y,Z)$ from the source to the target population.

%%%%%%%%%%%%%%%%%%%%%%%%
\subsection{Results}
Table~\ref{tab:dataresults-nhanes} reports point estimates, bootstrap standard errors based on $B=500$ resamples, $p$-values for testing whether the corresponding regression coefficients equal zero, and nominal 95\% confidence intervals.

\begin{table}[!ht]
  \centering
  \caption{Estimation and inference results for the NHANES target population.}\label{tab:dataresults-nhanes}
  \sisetup{
    table-format=-1.3,
    table-align-text-pre=false,
    table-align-text-post=false,
    input-open-uncertainty=(,
    input-close-uncertainty=),
    table-space-text-pre=(,
    table-space-text-post=),
  }
  \begin{threeparttable}
  \begin{tabular}{
    @{}
    c
    l
    c
    S[table-format=1.3]
    c
    c
    @{}
  }
    \toprule
    \multicolumn{1}{c}{\textbf{Method}} & \multicolumn{1}{c}{\textbf{Covariate}} & {\textbf{Estimate}} & {\textbf{SE}} & \multicolumn{1}{c}{\textbf{95\% CI}} & {\textbf{$p$-value}} \\
    \midrule
    \multirow{5}{*}{IW} 
    & $X$       & -0.014 & 0.039 & (-0.091,0.063) & 0.720 \\ 
    & energy    &  0.039 & 0.056 & (-0.072,0.150) & 0.488 \\ 
    & sex       &  0.146 & 0.103 & (-0.055,0.347) & 0.154 \\ 
    & age       &  0.142 & 0.043 & (0.057,0.226) & 0.001 \\ 
    \cmidrule{1-6}
    \multirow{5}{*}{IMP} 
    & $X$       & -0.015 & 0.039 & (-0.091,0.060) & 0.691 \\ 
    & energy    &  0.073 & 0.045 & (-0.014,0.161) & 0.101 \\ 
    & sex       &  0.204 & 0.089 & (0.030,0.378) & 0.021 \\ 
    & age       &  0.530 & 0.037 & (0.458,0.603) & $<0.001$ \\ 
    \cmidrule{1-6}
    \multirow{5}{*}{DRTL-CMC} 
    & $X$       & -0.272 & 0.184 & (-0.632,0.088) & 0.138 \\  
    & energy    &  0.057 & 0.049 & (-0.040,0.154) & 0.246 \\  
    & sex       &  0.287 & 0.123 & (0.045,0.529) & 0.020 \\  
    & age       &  0.575 & 0.070 & (0.437,0.712) & $<0.001$ \\  
    \bottomrule
  \end{tabular}
\begin{tablenotes}[flushleft]
\footnotesize
\item[] $X$ denotes average nightly sleep duration. SE denotes the bootstrap standard error.
\end{tablenotes}
\end{threeparttable}
\end{table}

As shown in Table~\ref{tab:dataresults-nhanes}, the estimated coefficient of sleep duration is negative under all three methods but is not statistically significant. DRTL-CMC gives a more negative point estimate (Estimate $=-0.272$, 95\% CI: $(-0.632,0.088)$; $p$-value $=0.138$) than IW and IMP, both of which yield estimates close to zero. The wide confidence interval for DRTL-CMC indicates substantial uncertainty, so the data provide limited evidence of an association between sleep duration and obesity in the target population.

Energy intake has small positive and non-significant coefficients under all three methods. The coefficient for sex is positive under all methods and is statistically significant under IMP and DRTL-CMC but not under IW. Age is positively associated with obesity under all three methods, although the estimated magnitude differs considerably between IW and the other two methods.

%%%%%%%%%%%%%%%%%%%%%%%%%%%%%%%%%%%%
\section{Discussion}\label{sec:discussion}
%%%%%%%%%%%%%%%%%%%%%%%%%%%%%%%%%%%%

This paper develops a doubly robust transfer learning framework for target generalized linear regression with completely missing covariates. By accommodating nonlinear target estimating equations, the proposed method extends existing linear-regression approaches while preserving double robustness and enabling valid large-sample inference.

Following \citet{zhao2026augmented}, identification relies on a sub-population shift assumption, which requires the conditional distribution of the missing covariates given the observed variables to be transportable across populations. This condition is not implied merely by the absence of the covariates from the target data and, because they are completely unobserved in the target population, cannot in general be verified from the observed target data alone. Its plausibility therefore depends on whether relevant source--target heterogeneity is adequately captured by the observed variables.

Our current implementation uses parametric working models for the density ratio and the conditional distribution of the missing covariates. More flexible nuisance estimators, including random forests and neural networks, could be incorporated to reduce reliance on parametric specifications. Extending the theoretical results to such estimators using cross-fitting and double machine learning techniques \citep{chernozhukov2018double} is an important direction for future research.

\endgroup

\paragraph*{Supplementary material}
The supplementary material contains all technical proofs, additional simulation results, and additional data analysis results. 

\bibliographystyle{abbrvnat}
\bibliography{ref}

\end{document}